\documentclass[pre,twocolumn,notitlepage,longbibliography,amsmath,amssymb,floats,superscriptaddress,nofootinbib,10pt]{revtex4-2}
\usepackage{graphicx,color}
\usepackage{amsmath}
\usepackage{amsfonts, amsbsy}
\usepackage{amssymb}
\usepackage{eqnarray}
\usepackage{bm}
\usepackage{blkarray}
\usepackage{mathtools}
\usepackage{psfrag}
\usepackage{caption}
\usepackage{subcaption}
\usepackage{multirow}
\usepackage{textcomp}
\usepackage{units}
\usepackage{lipsum}
\usepackage[title]{appendix}
\usepackage{soul}
\usepackage{titlesec}
\usepackage{times}
\usepackage{enumitem}
\usepackage{soul}
\usepackage[sort&compress]{natbib}
\usepackage[breaklinks,colorlinks = true,linkcolor = blue,urlcolor  = blue,citecolor = blue,anchorcolor = blue]{hyperref}
\usepackage{float}

\usepackage{xcolor}

\begin{document}
\title{Monopole magnetohydrodynamics on a plane: magnetosonic waves and dynamo instability}

\author{Debarghya Banerjee}
\email{banerjee@pks.mpg.de}
\affiliation{Institute for Theoretical Physics, University of Göttingen, Friedrich-Hund-Platz 1, 37077 Göttingen, Germany} 
\affiliation{Max Planck Institute for the Physics of Complex Systems, N\"othnitzer Stra\ss e 38, 01187, Dresden, Germany}

\author{Roderich Moessner}
\email{moessner@pks.mpg.de}
\affiliation{Max Planck Institute for the Physics of Complex Systems, N\"othnitzer Stra\ss e 38, 01187, Dresden, Germany}
\affiliation{W\"urzburg-Dresden Cluster of Excellence ct.qmat, 01187, Dresden, Germany}

\author{Piotr Sur\'owka}
\email{piotr.surowka@pwr.edu.pl}
\affiliation{Institute of Theoretical Physics, Wroc\l{}aw University of Science and Technology, 50-370 Wroc\l{}aw, Poland}

\date{\today}

\begin{abstract}
Condensed matter systems can host emergent `vacua' with particles, fields and  dimension different from that of the universe we inhabit. Motivated by the appearance of emergent gauge fields with both electric and magnetic charges, we consider the fate of electromagnetism in two dimensions in the presence of magnetic monopoles. 
We find that generically electromagnetic fields are damped due to resistive effects. However, we can still identify a magnetohydrodynamic regime. It exhibits magnetosonic waves which, unlike in $d=3$, are isotropic, to which there is a contribution from the odd viscosity coefficient present in the system. Further, we find 
a dynamo action, which unlike in $d=3$ requires compressibility of the flow. 
\end{abstract}

\maketitle 

\section{Introduction} 

Emergent gauge fields occur in several branches of topological condensed matter physics \cite{topobook}. This includes most prominently the quantum Hall effect \cite{PhysRevB.44.5246,Wen_IQHEedge}, in relation to both its bulk and its edge currents, and topological magnets called spin liquids \cite{anderson_resonating_1973,wen_quantum_2002,balents_spin_2010,Castelnovo2012,zhou_quantum_2017,Broholm_QSL,FieldGuide}. Increasingly complex types of emergent gauge fields are being discovered, for example tensor gauge theories~\cite{Pretko_2020} which appear in an entirely different type of setting such as  elasticity~\cite{RevModPhys.96.011001,Grosvenor:2021hkn}, folding paper~\cite{Manoj_2021} or jamming \cite{Namp_2022}.

It is then natural to ask not only what kind of new gauge theories exist, but also what phenomena they engender.  
This serves as a motivation for the present work, which studies the physics of electromagnetism beyond the familiar case of 3+1-dimensional Maxwell theory: we consider the properties of magnetohydrodynamics in two dimensions. 

Magnetohydrodynamics (MHD) ~\cite{Choudhuri1998,LifschitzV10} is the traditional description of the motion of charged particles and dynamics of magnetic fields. These equations are relevant particularly in understanding the co-evolution of matter and magnetic fields in the sun and solar winds~\cite{goldstein1995magnetohydrodynamic}, the earth's core and geomagnetism~\cite{roberts2013genesis,roberts1972magnetohydrodynamics}, and the motion of charged particles in the van Allen radiation belt leading to aurora-borealis~\cite{hones1986earth,blixt2006optical}. One classic aspect of MHD is the presence of Alfv\'en and magnetosonic waves with anisotropic dispersion relation~\cite{alfven1942existence,Choudhuri1998}.  Alfv\'en waves are observed in solar atmospheres~\cite{murawski2010linear,hollweg1978alfven}, cosmic rays~\cite{thomas2019cosmic,amato2018cosmic,Fermi1949}, and have a role to play in tokamak design used in nuclear fusion research~\cite{briguglio2000high,sharapov2002alfven,vlad1999dynamics}. Beyond these linear waves,  nonlinear aspects of MHD include the freezing of magnetic flux to the flow velocity field known as Alfv\'en's flux freezing theorem~\cite{hollweg1981alfven,eyink2011stochastic,Choudhuri1998}. The consequent transfer of energy between the velocity field and the electromagnetic field leads to a dynamo  action~\cite{krause2016mean,elsasser1956hydromagnetic,parker1963kinematical,parker1983hydrodynamics,priest2019magnetohydrodynamics,busse1978magnetohydrodynamics}, in which the kinetic energy of the fluid flow is converted to electromagnetic energy just like  a dynamo converts mechanical rotation of a dynamo into electricity. The presence of dynamo action is crucial in maintaining finite magnetic field in a diffusive or resistive environment. 

Restricting the dimensionality of the space inhabited by the gauge fields can lead to fundamental differences in their behavior, as already exemplified by the logarithmic rather than $1/r$ electrostatic potential in $d=2$ compared to $d=3$. In addition, one of the unique features of condensed matter systems is a possibility to realize not only emergent electric charges but also emergent magnetic charges, also known as magnetic monopoles.  This is in particular the case for  spin liquids  \cite{balents_spin_2010,Castelnovo2012,zhou_quantum_2017,Broholm_QSL,FieldGuide}, an interesting, yet somewhat elusive, class of systems in two spatial dimensions. These exotic  states lack long-range magnetic order even at very low temperatures. The effective field theory describing spin liquids can take the form of an emergent electrodynamics, sourced by monopole operators. Gauge fields in (2+1) dimensions are strongly coupled, which obstructs the application of perturbative methods. 

Here, using kinetic theory, we propose and analyze hydrodynamic equations describing the dynamics of magnetic monopoles in the presence of a two-dimensional gauge field at finite temperature. In particular, we study the linear magnetohydrodynamic regime and show the presence of elementary excitations which are analogous to the magnetosonic waves in magnetohydrodynamics. However, unlike the natural anisotropy present in magnetosonic waves in (3+1) dimensional setup these two-dimensional waves are isotropic. We further show that there can be an incarnation of the dynamo instability, different from the usual MHD in that compressibility of the fluid is an essential ingredient. It is important to note, much as in other derivations of hydrodynamics from kinetic theory, that the resulting hydrodynamic equations remain valid even in strongly coupled systems. In this sense, kinetic theory serves as a useful tool for determining the form of these equations, whose applicability extends well beyond systems with weakly coupled microscopic constituents.

\section{Emergent electrodynamics on a plane} 

Equations of motion of electromagnetic fields in (2+1) dimensions, in the presence of both electric and magnetic charges, can be presented in the following form:
\begin{subequations}\label{eq:Maxwelle}
\begin{equation} 
\nabla \cdot {\bm E} = 2\pi \rho_e,
\end{equation}
\begin{equation} 
({\bm \epsilon} \cdot \nabla) \cdot {\bm E} = \frac{1}{c} \frac{\partial B}{\partial t}, 
\end{equation}
\begin{equation} 
{\bm \epsilon} \cdot \nabla B = \frac{1}{c} \frac{\partial {\bm E}}{\partial t} + \frac{2\pi}{c} {\bm J}_e,
\end{equation}
\end{subequations}
where ${\bm \epsilon} $ is the Levi-Civita symbol. The degrees of freedom are a two-component electric field and a (pseudo)scalar magnetic field. This is to be contrasted with the three-dimensional case, in which the electric and magnetic fields are three-dimensional vectors and pseudovectors, respectively. Both Maxwell's equations in (3+1) dimensions and MHD equations in the presence of magnetic monopoles are invariant under the duality transformation \cite{coceal_duality-invariant_1996}
\begin{equation} \label{eq:dictionary}
\rho_e \rightarrow \rho_m, \quad \rho_m \rightarrow -\rho_e, \quad {\bm E}_{\text{3d}} \rightarrow {\bm B}_{\text{3d}}, \quad {\bm B}_{\text{3d}} \rightarrow -{\bm E_{\text{3d}}}.
\end{equation}
Inspired by this, one may apply an analogous transformation in two dimensions to arrive at
\begin{subequations}\label{eq:Maxwellm}
\begin{equation} 
\nabla \cdot {\bm B}_m = 2\pi \rho_m,
\end{equation}
\begin{equation} 
({\bm \epsilon} \cdot \nabla) \cdot {\bm B}_m = -\frac{1}{c} \frac{\partial E_m}{\partial t}, 
\end{equation}
\begin{equation} 
-{\bm \epsilon} \cdot \nabla E_m = \frac{1}{c} \frac{\partial {\bm B}_m  }{\partial t} + \frac{2\pi}{c} {\bm J}_m.
\end{equation}
\end{subequations}
Note that, in contrast to three dimensions, $\rho_e$ and $\rho_m$ are not simultaneously present in Maxwell's equations, which means that electric and magnetic charges constitute two decoupled sectors of the duality, and as such, our results apply to both.  Additionally, the presence of duality is not correlated with the specific physical details of the system, whether it contains fields and charges from two sectors of the theory or only one.  We see that in the dual formulation, the magnetic field becomes vectorial and electric field becomes a scalar. Because the magnetic and electric charges are decoupled dual formulations are completely equivalent. In the original formulation we have electric charges and in the dual formulation we have magnetic charges, which we refer to as monopoles.  As a result one can formulate magnetohydrodynamic evolution either in terms of the original variables with vector electric field and scalar magnetic field or using the dual fields. The duality maps original MHD equations into equivalent electrohydrodynamic equations. In what follows we find it convenient to use the dual formulation, still refering to it as MHD. In standard terminology, a duality represents the relationship between two theories that are equivalent through a defined dictionary \cite{carl_dualities_2019}. In our case, this dictionary is explicitly specified by a mapping that transforms Eqs. \eqref{eq:Maxwelle} into Eqs. \eqref{eq:Maxwellm}. We emphasize that this mapping does not involve a transformation between strong and weak coupling regimes.

\section{MHD with monopoles} 

We now focus on two-dimensional electromagnetism with no electric charge and only magnetic monopoles. These emergent charges are tied to a two-dimensional gauge field. The effective Lorentz force (force density ${\bm f}$) experienced by these emergent monopoles is given by:
\begin{align}
    {\bm f} = \rho_m {\bm B}_m + \frac{({\bm \epsilon} \cdot {\bm J}_m) E_m}{c},
    \label{eq:lorentz}
\end{align}
where $\rho_m$ is the density of magnetic monopoles, ${\bm B}_m$ is the magnetic field experienced by the monopoles, $E_m$ is the pseudo-scalar electric field experienced by the magnetic monopoles, ${\bm J}_m$ is the current density due to the motion of monopoles, $c$ is the speed of light in the emergent two-dimensional gauge field. The connection of these coarse-grained variables to microscopic variables is derived in  Appendix \eqref{app:A} below. Using the expression of Lorentz force from Eq.~\ref{eq:lorentz} we can obtain continuum equations. 

We assume a steady-state behavior of the current due to magnetic monopoles i.e. $d {\bm J}_m/ dt \rightarrow 0$ which gives us the effective Ohm's law:
\begin{align}
    {\bm J}_m = \frac{\chi \rho}{2 \pi} \left( {\bm B}_m + \frac{1}{c} E_m {\bm \epsilon} \cdot {\bm v} \right),
    \label{eq:ohms}
\end{align}
where $\rho$ is the mass density, $\chi$ is a parameter characterising the effective conductivity of the monopoles, and ${\bm v}$ is the velocity field of the monopole-fluid. The form of ${\bm J}_m$ needs to be considered along with Maxwell's equations:
\begin{subequations}
\begin{equation}
     \frac{\partial E_m}{\partial t} = -c \nabla \times {\bm B}_m, 
     \end{equation}
\begin{equation}
     {\bm J}_m \approx -\frac{c}{2 \pi} {\bm \epsilon} \cdot \nabla E_m.
    \label{eq:maxwells}
\end{equation}
\end{subequations}
In the above form of Maxwell's equations we have assumed that $\rho_m = 0$, i.e. there is no major charge separation and the system remains neutral at the level of coarse-graining. Further,  
\begin{align}
    \frac{1}{c} \left| \frac{\partial {\bm B}_m}{\partial t} \right| \frac{1}{ |{\bm \epsilon} \cdot \nabla E_m|} \approx \frac{|{\bm B}_m|}{E_m} \frac{l}{c t} \rightarrow 0,
    \label{eq:mhd_limit}
\end{align}
which implies that the length scale ($l$) of fluctuations in $E_m$ is small, the dynamics of ${\bm B}_m$ is slow ($t$ is large), and the overall  amplitude of fluctuations in ${\bm B}_m$ is smaller than the amplitude of fluctuations in $E_m$. In these units, $E_m$ and ${\bm B}_m$ have the same dimension and therefore the ratio mentioned in Eq.~\ref{eq:mhd_limit} is dimensionless. 

In this limit we can take a curl of the Ohm's law and use Maxwell's equations from Eq.~\ref{eq:maxwells} to obtain an evolution equation for $E_m$. If we consider the evolution of momentum we obtain an effective Navier-Stokes equation, and considering an effective conservation of mass gives a continuity equation for density. Therefore, the energy due to the field fluctuations is present mainly in the $E_m$ field rather than the ${\bm B}_m$ field. This is also reflected in the fact that we obtain a hydrodynamic equation for $E_m$ and not for ${\bm B}_m$. Having established the evolution equations for the gauge fields we supplement them with the evolution equations for the fluid, that take the form of conservation laws
\begin{subequations}
\begin{equation}
         \frac{\partial \rho}{\partial t} = -\nabla \cdot {\bm j},
\end{equation}
\begin{equation}
\frac{\partial (\rho {\bm v})}{\partial t} = -\nabla \cdot {\bm \tau} + {\bm f},
\end{equation}
\end{subequations}
where ${\bm j}$, ${\bm \tau}$ correspond to currents of density and momentum and ${\bm f}$ is the force density defined in Eq.~\ref{eq:lorentz}. In order to get a closed set of MHD equations we perform the gradient expansion of these quantities in terms of fluid and gauge degrees of freedom 
\begin{subequations}
\label{eq:hydro_eq}
\begin{equation}
     \frac{\partial \rho}{\partial t} = -\rho_0 \nabla \cdot {\bm v},    \end{equation}
\begin{equation}\frac{\partial {\bm v}}{\partial t} = \frac{\eta}{\rho_0} \nabla^2 {\bm v} + \frac{\eta^o}{\rho_0} \nabla^2 ({\bm \epsilon} \cdot {\bm v}) - \frac{a}{\rho_0} \nabla \rho - \frac{1}{\rho_0 \pi} E_m \nabla E_m,    \end{equation}
\begin{equation} \frac{\partial E_m}{\partial t} = D \nabla^2 E_m - \frac{1}{2 \pi} E_m \nabla \cdot {\bm v},
\end{equation}
\end{subequations}
where we have ignored the advective nonlinearity in the fluid and density evolution equation and furthermore we have assumed that the pressure ($p$) in the fluid equation can be approximated as $p = p_0 + a (\rho - \rho_0) + \ldots$ to leading order in density fluctuations (see App. \eqref{app:A} for details). The parameter $\eta$ is the shear viscosity, $\eta^o$ is the odd viscosity~\cite{Avron1995,Avron1998}, $D$ is the diffusivity of the field $E_m$ arising due to resistive effects, and $\rho_0 = \langle \rho \rangle $. The presence of odd viscosity is crucial in the hydrodynamics of two-dimensional chiral fluids, such as our planar MHD. Examples of other fluids with odd viscosity in two dimensions include hydrodynamics of classical and quantum states in the presence of magnetic fields and fluids of rotating objects that are most notably realized in active matter ~\cite{Banerjee2017,Ganeshan2017,Souslov2019,Banerjee2022,soni2019odd,Banerjee_PhysRevLett.126.138001,Lier_PhysRevE.108.L023101,Hosaka_PhysRevLett.131.178303}. 

\section{Hydrodynamic waves} 

For weak fluctuations in $E_m$ around $\langle E_m \rangle = 0$, the Lorentz force drops out from the evolution equation for velocity as it is nonlinear in the fluctuation in $E_m$. The evolution of $E_m$ then reduces to a linear diffusive equation. However, if $\langle E_m \rangle = E_0$ the Lorentz force term in the velocity evolution equation becomes $E_0 \nabla E_m$ and the evolution equation for $E_m$ has a linear term $E_0 \nabla \cdot {\bm v}$. We can assume solutions of the form $\exp{i( \omega t - {\bm k} \cdot {\bm x})}$, where $\omega$ is the angular frequency of the wave, ${\bm k}$ is the wavevector, $t$ is the time, and ${\bm x}$ is the spatial coordinate. We find that these waves obey the dispersion relation:
\begin{align}
    & \omega = \pm k \sqrt{a + \frac{E_0^2}{2 \pi^2 \rho_0} + \frac{{\eta^o}^2 k^2}{\rho_0^2}},
\end{align}
where $k = |{\bm k}|$. In order to derive the above dispersion relation we have ignored the effect of diffusion and shear-viscosity. The parameter $a$ effectively gives us the sound speed in the system and the odd viscosity correction appears as expected  {from hydrodynamic theory} modifying the wave speed for large values of $k$. In the limit of weak coupling to density fluctuations i.e., $a \ll E_0^2 / \rho_0$, we obtain waves with phase velocity proportional to $E_0$. This is analogous to the magnetosonic waves observed in the usual MHD picture with the key difference that magnetosonic waves have a velocity proportional to $B_0 f(\theta)$, where $B_0$ is the mean magnetic field, $\theta$ is a polar angle, and $f(\theta)$ is a function of the polar angle. In the usual MHD picture Alfv\'en and magnetosonic waves are calculated as a fluctuation about a mean magnetic field and since that is a vector this implies choosing a direction along with a mean magnetic field magnitude. Therefore, the polar angle $\theta$ appears in the expression for Alfv\'en and megnetosonic wave velocity. However, in the two dimensional monopolic MHD, the mean electric field is a pseudoscalar and the magnetosonic velocity is isotropic.

\begin{figure}
    \centering
    \includegraphics[width=\linewidth]{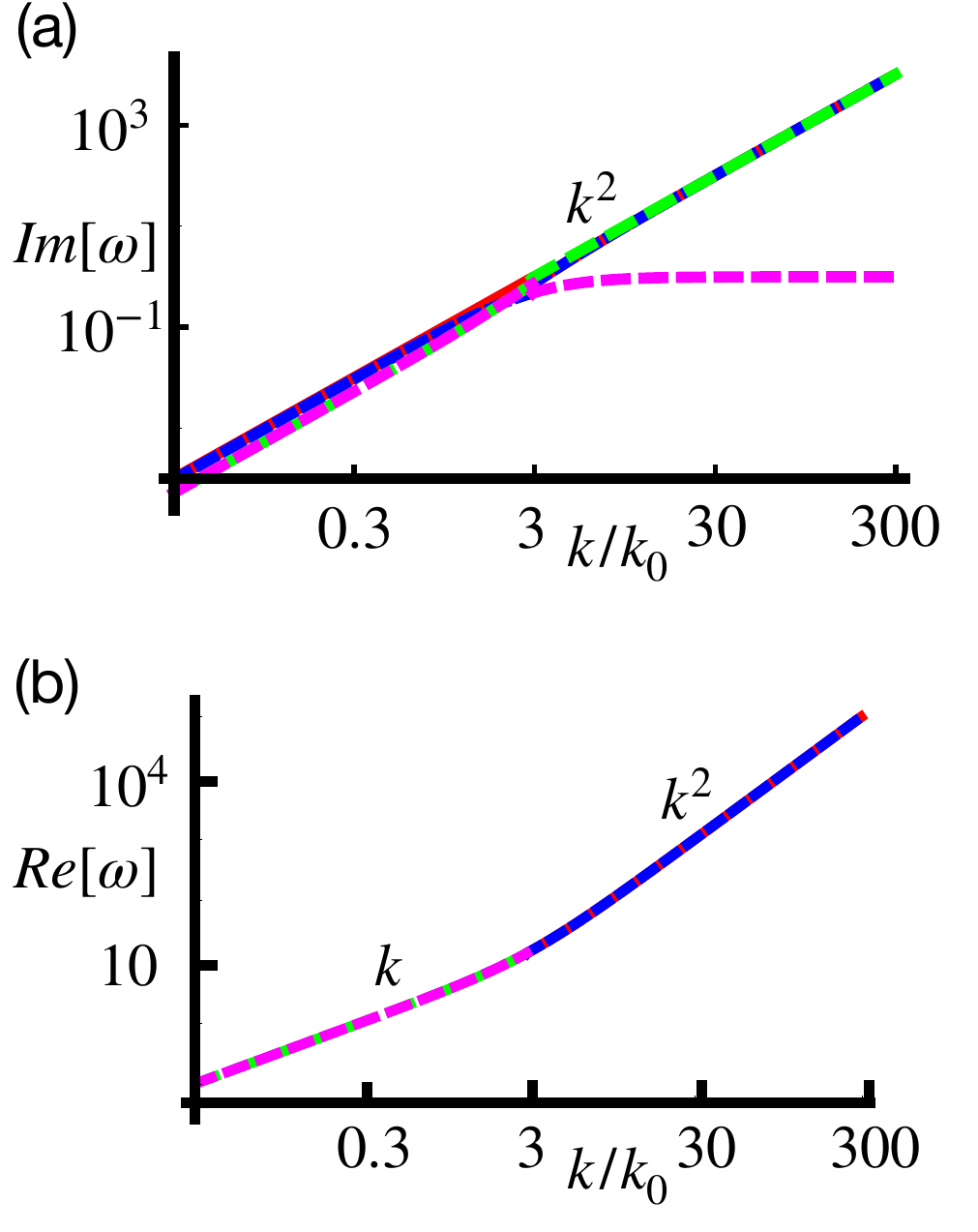}
    \caption{\textbf{Dispersion relation}: Numerical solutions for the dispersion relations. In (a) we plot the imaginary part of $\omega$ versus wavenumber rescaled by a characteristic wavenumber $k_0$, where, $k_0 = E_0/ \pi \eta^o$ and in (b) we plot the real part of $\omega$. The different colors correspond to the four different eigenvalues.}
    \label{fig:dispersion_relation}
\end{figure}

In the presence of dissipative and diffusive terms we compute their effects perturbatively in powers of wavenumber. This gives us:
\begin{align}
    & \omega_1  = i \frac{2 a D \rho_0 \pi^2 }{E_0^2 + 2 a \pi^2 \rho_0} k^2 + \mathcal{O} (k^4), \nonumber \\
    & \omega_2 = i \frac{\eta}{\rho_0} k^2 + \mathcal{O} (k^4),
\end{align}
and damped waves given by :
\begin{widetext}
\begin{align}
     \omega_{3,4} = & \pm  k \sqrt{\frac{E_0^2 + 2 a \pi^2 \rho_0}{2 \pi^2 \rho_0}} + i k^2 \frac{2 a \eta \pi^2 \rho_0 + E_0^2 (\eta + D \rho_0)}{2 \rho_0 (E_0^2+ 2 a \pi^2 \rho_0)} \nonumber \\
    & \pm k^3 \frac{\pi \rho_0 (4 a^2 (\eta^2 - 4 {\eta^o}^2) \pi^4 \rho_0^2 + E_0^4 (\eta^2 - 4 {\eta^o}^2 - 2 D \eta \rho_0 + D^2 \rho_0^2) + 4 a E_0^2 \pi^2 \rho_0 (\eta^2 - 4 {\eta^o}^2 - D \eta \rho_0 + 2 D^2 \rho_0^2))}{4 \sqrt{2} (\rho_0 (E_0^2 + 2 a \pi^2 \rho_0))^{5/2}} + \mathcal{O} (k^4).
\end{align}
\end{widetext}
In the above expressions $\omega_{1,2,3,4}$ are the expressions for frequency $\omega$. We further solve the Eqs.~\ref{eq:hydro_eq} numerically and show the behavior of the real and imaginary parts of $\omega$ in Fig.~\ref{fig:dispersion_relation} (b) and (a) respectively. We clearly see the $\omega \sim k$ and $\omega \sim k^2$ regimes in the real parts of $\omega$. For the imaginary part we see the $\omega \sim k^2$ regime. One of the imaginary parts of eigenvalues exhibit saturation for large $k$ although for small $k$ it retains the diffusive nature.

We now consider the nonlinear term $-E_m \nabla \cdot {\bm v}$ in the evolution equation of $E_m$. While at the linear level the equation is a diffusion equation, the nonlinear term yields a coupling between the velocity field and $E_m$. In traditional MHD this type of nonlinear term leads to what is called the dynamo action. In fact this coupling is responsible for sustaining a magnetic field for long times in a resistive environment.
\begin{figure}
    \centering
    \includegraphics{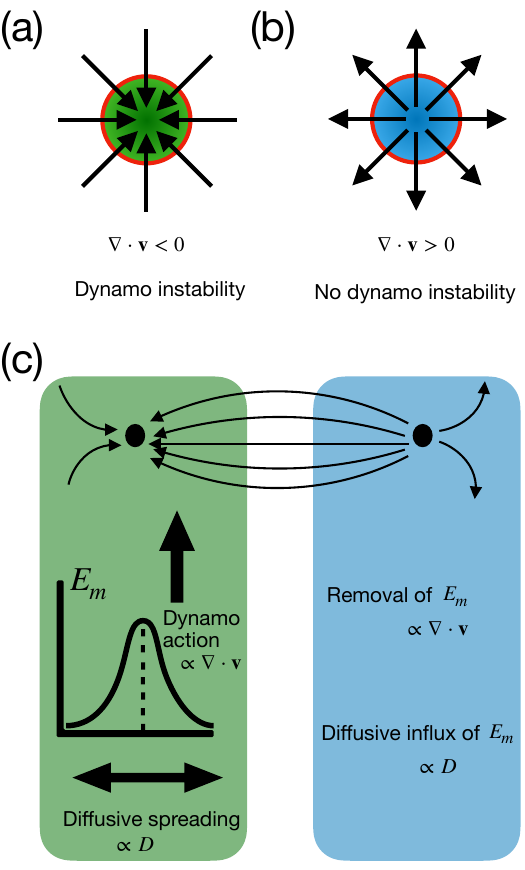}
    \caption{The arrows correspond to the velocity field (a) In regions of $\nabla \cdot {\bm v} < 0$ there are a sink and dynamo instability, while in (b) regions of $\nabla \cdot {\bm v} > 0$ there are a source and no dynamo instability. (c) Schematic flow with both source and sink, and the corresponding effect in terms of dynamo action and diffusion.}
    \label{fig:dynamo}
\end{figure}

In our two-dimensional set up the dynamo action is connected to the compressibility of the fluid unlike the usual MHD system where dynamo action also takes place in an incompressible fluid, where $\nabla \cdot {\bm v} \equiv 0$. In regions of negative $\nabla \cdot {\bm v} < 0$,  there is an exponential growth of the $E_m$ field leading to the dynamo instability (see Fig. \ref{fig:dynamo}). 
This is usually unstable because of the piling up of mass in these regions. In our case there is a piling up of the $E_m$, leading to a growth of the local electromagnetic energy at the cost of the kinetic energy of flow. The diffusivity then acts as a mechanism to spread out this concentrated electromagnetic energy.

Beyond the MHD approximation, i.e.\ when $({\bm B}_m l)/(c t E_m) \sim 1$ (see Eq.~\ref{eq:mhd_limit}), we have Maxwell's equation $- {{\bm \epsilon} \cdot \nabla} E_m = \frac{1}{c} \frac{\partial {\bm B}_m}{\partial t} + \frac{2 \pi}{c} {\bm J}_m$. This gives us:
\begin{align}
    \frac{\partial E_m}{\partial t} = D \nabla^2 E_m - \frac{1}{2 \pi} E_m \nabla \cdot {\bm v} - \frac{1}{2 \pi c}\frac{\partial^2 E_m}{\partial t^2}.
    \label{eq:not_mhd}
\end{align}
The above equation gives rise to damped oscillatory solutions for $E_m$.

\section{Discussion and outlook} 
We have developed magnetohydrodynamics of a monopole fluid on a plane. We have employed the electromagnetic duality that changes the role of electric and magnetic fields as well as electric and magnetic charges. We have obtained linear waves which have a wave velocity proportional to the mean electric field $E_0$. These wave solutions are similar to the magnetosonic waves observed in the usual MHD picture, however, they are isotropic as opposed to the anisotropic magnetosonic waves in MHD. In addition we have also shown that odd viscosity plays an important role in the dispersion relation. This is a consequence of the symmetry pattern in two dimensions in the presence of magnetic field (see also \cite{Avron1995,Avron1998,Abanov2018,Lingam2014,Lingam2015,Lingam2015hall,Lingam2020,Lucas2014,Passot2007,Passot2017}). If we consider the nonlinear effects in the electric field we find a dynamo instability arising in certain regimes of compression. This is different from the usual dynamo action observed in MHD where compressibility of the flow is not a prerequisite for the dynamo action. 

What about physical realisations of such physics? In condensed matter physics, many quantities are tunable: besides the obviously variable temperature, the geometric arrangement of degrees of freedom, including its  dimensionality, can be varied. The emergent low-energy degrees of freedom can in turn take the form of an emergent gauge field, while the quasiparticles may carry gauge charges of its electric or magnetic~\cite{senthil_deconfined_2004,metlitski_monopoles_2008,hermele_non-abelian_2009} (or both) sectors. 
They may at the same time carry charges of Maxwell electromagnetism, such as the magnetic monopoles in spin ice in three dimensions \cite{Castelnovo2008}. Quite prominently, their cousins in two-dimensional 
artificial spin ice (ASI), a class of engineered magnetic nanostructures, have also been studied \cite{ladak_direct_2010,otsuka_scaling_2014,gilbert_emergent_2014,farhan_emergent_2019,miao_two-dimensional_2020,takatsu_universal_2021,goryca_field-induced_2021,duarte_magnetic_2024}. 

Indeed, in the latter, the Maxwell magnetic charge exhibits a three-dimensional Coulomb $1/r$ law, despite the fact that emergent fields reside in two dimensions.
Nevertheless, at the macroscopic level, the specific details of these interactions are encapsulated within the values of the transport coefficients. As long as the emergent degrees of freedom and symmetries remain the same, the form of the hydrodynamic equations will not change. Advances in transport measurements in these materials may enable us to verify the properties of two-dimensional magnetohydrodynamics, as discussed in this paper. 

Most immediately, the goal is to find a quantum spin liquid in which the emergent gauge field is two-dimensional. However, in two dimensions, pure U(1) spin liquids are confining at zero temperature. On one hand, one could generalize our construction to include other gauge groups. 
At the same time, it is also conceivable that there exists a regime in which our finite-temperature treatment is applicable. 
Establishing a connection between the low-energy magnetohydrodynamic regime and the microscopic models of quantum spin liquids requires an understanding of the length and time scales underpinning a hydrodynamic regime, and how they might be realised in a quantum spin liquid. 
For completeness, we note that intrinsic momentum conservation is already broken due to umklapp processes, and there is no gauge symmetry. However, these two symmetries are fundamental to the hydrodynamic equations used here. While some arguments on the validity of hydrodynamic regime have been presented in Ref.~\cite{Bulchandani2021}, the precise understanding of when hydrodynamics and magnetohydrodynamics work in quantum spin liquids requires a more detailed investigation.

Finally our model is phenomenological. A first-principles derivation of MHD equations is done by means of an unusual type of symmetry dubbed a one-form symmetry. It would be beneficial to formulate and extend our phenomenological model based on principles of symmetry \cite{grozdanov_generalized_2017,armas_magnetohydrodynamics_2019,armas_one-form_2020,armas_approximate_2024}.

\section{Acknowledgements} 
This work was supported in part by
the Deutsche Forschungsgemeinschaft under Grant No. SFB
1143 (Project-ID No. 247310070),
the Deutsche Forschungsgemeinschaft under cluster of excellence
ct.qmat (EXC 2147, Project-ID No. 390858490), and the Polish National Science Centre (NCN) Sonata Bis grant 2019/34/E/ST3/00405.

\bibliography{bibliography.bib}

\section{Appendix A : Kinetic derivation of the dynamics of magnetic monopoles} \label{app:A}

In this section we derive the general equations that give the time evolution in the dynamics of magnetic monopoles. We choose a kinetic theory model, in analogy with the standard treatment of three-dimensional plasmas \cite{nicholson_introduction_1983}. To begin with we need to consider the forces acting on the charges in the two dimensional gauge field. Without loss of generality we can divide the force involved into four parts : The first part is the emergent electric field (${\bm E}_e$) acting on the electric charge density ($\rho_e$), the second part consists of the emergent magnetic field ($B_e$) acting on the emergent electric current density (${\bm J}_e$), the third part is due to the emergent magnetic field (${\bm B}_m$) acting on the magnetic charge density ($\rho_m$), and the fourth part is composed of emergent electric field ($E_m$) acting on the emergent magnetic current density (${\bm J}_m$). Altogether the force density is given by:
\begin{align}
    {\bm f} = \rho_e {\bm E}_e + \frac{({\bm \epsilon}\cdot {\bm J}_e) B_e}{c} + \rho_m {\bm B}_m + \frac{({\bm \epsilon}\cdot {\bm J}_m) E_m}{c},
\end{align}
in the above description $c$ is the speed of electromagnetic waves in the medium and ${\bm \epsilon}$ is the two-dimensional Levi-Civita tensor.  If we now consider a system with no electric charge or current but finite magnetic charge and current, then $\rho_e = 0$ and ${\bm J}_e = 0$. 

The positive and negative charged magnetic monopoles experience the force described in the above paragraph and undergo collisions. Since positive and negative charges attract each other the dominant collisions are give by the exchange of momentum between positive and negative charged particles. Therefore, the dynamics of the particles is given by:
\begin{subequations}
\begin{equation}
\begin{split}
     m n_{+} \frac{d}{d t} {\bm v}_{+} &= n_{+} g \left( {\bm B}_m + \frac{({\bm \epsilon}\cdot {\bm v}_{+}) E_m}{c}  \right)  \\ &+ n_{+} m C_f ({\bm v}_{-} - {\bm v}_{+}), 
     \end{split}
     \end{equation}
\begin{equation}
\begin{split}
     m n_{-} \frac{d}{d t} {\bm v}_{-}& = -n_{-} g \left( {\bm B}_m + \frac{({\bm \epsilon}\cdot {\bm v}_{-}) E_m}{c}  \right) \\ & + n_{-} m C_f ({\bm v}_{+} - {\bm v}_{-}),
     \end{split}
     \end{equation}
\end{subequations}
where, $g$ is the unit magnetic charge of a particle, $m$ is the mass of the particle (we assume the mass of the positive charge $m_{+}$ and the mass of the negative charge $m_{-}$ are same unlike in usual electromagnetism where the positive charge is much more massive than the negative charge), $n_{+}$ is the number density of the positive charge, $n_{-}$ is the number density of the negative charge, ${\bm v}_{+}$ is the velocity of the positive charge, ${\bm v}_{-}$ is the velocity of the negative charge, and $C_f$ is collision frequency of positive and negative magnetic charges. From the above microscopic description it is possible to define coarse-grained variables as:
\begin{subequations}
\begin{equation}
 {\bm J}_m = g (n_{+} {\bm v}_{+} - n_{-} {\bm v}_{-}),
\end{equation}
\begin{equation}
 \rho {\bm v} = m (n_{+} {\bm v}_{+} + n_{-} {\bm v}_{-}), 
\end{equation}
\begin{equation}
     \rho = m(n_{+} + n_{-}),  
\end{equation}
\begin{equation}
     \rho_m = g(n_{+} - n_{-}).
\end{equation}
\end{subequations}
This gives $n_{+} = \rho/(2m) + \rho_m/(2g)$ and $n_{-} = \rho/(2m) - \rho_m/(2g)$. In terms of the continuum variables we have the following equations:
\begin{subequations}
\begin{equation}
    \frac{d}{d t} (\rho {\bm v})  = \rho_m {\bm B}_m + \frac{2}{c} E_m ( {\bm \epsilon} \cdot {\bm J}_m), 
\end{equation}
\begin{equation}
     \frac{d}{d t} {\bm J}_m  = \frac{g^2}{m^2} \rho \left( {\bm B}_m + \frac{1}{c} E_m {\bm \epsilon} \cdot {\bm v} \right)- 2 C_f {\bm J}_m. 
\end{equation}
\end{subequations}
Here, the last term in the second equation involves an approximation -- what we actually get in the term involving collision frequency is $C_f g(n_{+} {\bm v}_{+} -n_{-} {\bm v}_{-}) + C_f g (n_{+} {\bm v}_{-} - n_{-} {\bm v}_{+})$. The second part of this term is taken to be ${\bm J}_m$ as an approximation only and hence the collision term has $2 {\bm J}_m$. The first equation gives us the force acting on the fluid and adds to the usual Navier-Stokes equation for fluid dynamics. Steady-state current gives us the effective Ohm's law for this monopole current:
\begin{align}
    {\bm J}_m = \frac{g^2 \rho}{2 m^2 C_f} \left( {\bm B}_m + \frac{1}{c} E_m {\bm \epsilon} \cdot {\bm v} \right).
\end{align}
At this stage it is convenient to define parameter $\chi \equiv g^2 \pi /m^2 C_f$ for use later. The first equation obtained from the kinetic description gives the mechanical force and ignoring advective nonlinearities we have:
\begin{equation}
\begin{split}
    \frac{\partial {\bm v}}{\partial t}& = \frac{\eta}{\rho} \nabla^2 {\bm v} + \frac{\eta^o}{\rho} \nabla^2 ({\bm \epsilon} \cdot {\bm v}) - \frac{\nabla p}{\rho} \\& + \frac{\rho_m}{\rho}{\bm B}_m + \frac{2}{c \rho} E_m ( {\bm \epsilon} \cdot {\bm J}_m),
    \end{split}
\end{equation}
where $\eta$ is the shear viscosity and $\eta^o$ is the odd viscosity, $p$ is the hydrostatic pressure. The mass density and the monopole density obeys continuity equations given by:
\begin{subequations}
\begin{equation}
    \frac{\partial \rho}{\partial t} = -\nabla \cdot (\rho { \bm v}), \end{equation}
\begin{equation}
    \frac{\partial \rho_m}{\partial t} = - \nabla \cdot {\bm J}_m.
\end{equation}
\end{subequations}
In addition we have the two dimensional Maxwell's equations which are fundamental to the gauge field:
\begin{subequations}
\begin{equation}
    \nabla \cdot {\bm B}_m = 2 \pi \rho_m, \end{equation}
\begin{equation}
     \nabla \times {\bm B}_m = -\frac{1}{c} \frac{\partial E_m}{\partial t}, \end{equation}
\begin{equation}
     - {{\bm \epsilon} \cdot \nabla} E_m = \frac{1}{c} \frac{\partial {\bm B}_m}{\partial t} + \frac{2 \pi}{c} {\bm J}_m.
\end{equation}
\end{subequations}
It is to be noted that the last Maxwell's equation and the continuity of monopole density are effectively the same equations. This can be seen by taking a divergence of the last equation above and using Gauss's law. Now, we have the complete set of equations obeyed by a two-dimensional fluid of emergent magnetic monopoles. The presence of finite $\rho_m$ i.e. when $n_{+} \neq n_{-}$ is useful in the study of plasma oscillations like physics. In electronics the dynamics is determined by the motion of electrons and static positive charge. This gives us the gapped longitudinal plasma oscillations. In the presence of odd viscosity the oscillations do not remain exclusively longitudinal and transverse modes develop. 

In the limit where $\rho_m = 0$ the equations reduce to:
\begin{subequations}
\begin{equation}
    \frac{\partial {\bm v}}{\partial t} = \frac{\eta}{\rho} \nabla^2 {\bm v} + \frac{\eta^o}{\rho} \nabla^2 ({\bm \epsilon} \cdot {\bm v}) - \frac{1}{\rho}\nabla p + \frac{2}{c \rho} E_m ( {\bm \epsilon} \cdot {\bm J}_m), \end{equation}
\begin{equation} \frac{\partial \rho}{\partial t} = -\nabla \cdot (\rho { \bm v}), \end{equation}
\begin{equation} 
\nabla \times {\bm B}_m = -\frac{1}{c} \frac{\partial E_m}{\partial t},  \end{equation}
\begin{equation}
{\bm J}_m = \frac{\chi \rho}{2 \pi} \left( {\bm B}_m + \frac{1}{c} E_m {\bm \epsilon} \cdot {\bm v} \right), 
\end{equation}
\begin{equation}
- {{\bm \epsilon} \cdot \nabla} E_m = \frac{1}{c} \frac{\partial {\bm B}_m}{\partial t} + \frac{2 \pi}{c} {\bm J}_m .
\end{equation}
\end{subequations}
Now, let's consider the linearized versions of these equations. By linearizing around a mean density and a mean zero electric field (with $\rho_0 \equiv \langle \rho \rangle$, $\langle E_m \rangle = 0$, and $\langle {\bm B}_m \rangle \neq 0$), the electromagnetic sector becomes decoupled from the momentum equations. As a result, we are effectively left with exponentially decaying electromagnetic waves and sound waves, which include corrections for odd viscosity in the mechanical sector. In the other case if we linearize about finite mean electric field i.e. $\langle E_m \rangle = E_0 \neq 0$ we obtain coupled equations. We further consider the pressure to be a function of density and write it as $p = \Sigma_i b_i (\rho -\rho_0)^i$ and retain the leading order as $p = p_0 + a (\rho - \rho_0) + \ldots$. 
\begin{subequations}
\begin{equation}
     \frac{\partial {\bm B}_m}{\partial t} = -\chi {\bm B}_m - \frac{1}{c} \chi  E_0 {\bm \epsilon} \cdot {\bm v} 
     \end{equation}
\begin{equation} 
\rho_0 \frac{\partial {\bm v}}{\partial t} = \eta \nabla^2 {\bm v} + \eta^o \nabla^2 ({\bm \epsilon} \cdot {\bm v}) - a \nabla \rho + \frac{2}{c} \chi E_0 \left( {\bm \epsilon} \cdot   {\bm B}_m + \frac{1}{c} E_0  {\bm v} \right), \end{equation}
\begin{equation} \frac{\partial \rho}{\partial t} = - \rho_0 \nabla \cdot {\bm v}.
\end{equation}
\end{subequations}
If we now consider an incompressible fluid i.e. $\nabla \cdot {\bm v} = 0$ then the density fluctuations are zero and the velocity field can be reduced to the vortical structures given by the $\nabla \times {\bm v}$ which implies that the vorticity evolution is affected by the divergence of magnetic field and the vorticity affects the evolution of the divergence of the magnetic field. But the divergence of the magnetic field is zero because of the absence of the magnetic charge ($\rho_m = 0$). Therefore, we have a simple situation of exponentially damped magnetic field and the velocity field having an exponentially growing vorticity due to the presence of the mean electric field.

\section{Appendix B : Hydrodynamic regime and magnetosonic waves} \label{app:B}

To obtain the hydrodynamic regime we need to approximate Maxwell's equations:
\begin{subequations}
\begin{equation}
     \frac{\partial E_m}{\partial t} = -c \nabla \times {\bm B}_m, 
     \end{equation}
\begin{equation}  
{\bm J}_m \approx -\frac{c}{2 \pi} {\bm \epsilon} \cdot \nabla E_m.
\end{equation}
\end{subequations}
In the above approximation we have considered a slow evolution of the magnetic field. A scaling argument can be constructed to clarify the regime as:
\begin{align}
    \frac{1}{c} \left| \frac{\partial {\bm B}_m}{\partial t} \right| \frac{1} {|{\bm \epsilon} \cdot \nabla E_m|} \approx \frac{|{\bm B}_m|}{E_m} \frac{l}{c t},
\end{align}
where $l$ is the typical length scale over which the electric field varies and $t$ is the typical time scale of magnetic field fluctuations. This ratio is small and the approximation of the hydrodynamic regime works when the time scale involved with magnetic field evolution is large and the length scale over which electric field fluctuations vary is small i.e. $l/t \ll c$. As we will see later that the the equations of hydrodynamics involve an equation for $E_m$ this means from a hydrodynamic perspective that $E_m$ fluctuations are the slow variables and not the fluctuations in ${\bm B}_m$ this is also reflected in the scaling argument above i.e. $E_m \gg |{\bm B}_m|$. Ohm's Law for this system is:
\begin{align}
    {\bm J}_m = \frac{\chi \rho}{2 \pi} \left( {\bm B}_m + \frac{1}{c} E_m {\bm \epsilon} \cdot {\bm v} \right).
\end{align}
Taking a curl and using Maxwell's equations (with the approximations described above) we get:
\begin{subequations}
\begin{equation}
     \nabla \times {\bm J}_m = -\frac{\chi \rho}{c} \partial_t E_m - \frac{\chi \rho}{2 \pi c} E_m \nabla \cdot {\bm v},      \end{equation}
\begin{equation}
     -\frac{c}{2 \pi} \nabla^2 E_m = -\frac{\chi \rho}{c} \partial_t E_m - \frac{\chi \rho}{2 \pi c} E_m \nabla \cdot {\bm v}.
\end{equation}
\end{subequations}
Here we have the equations describing the electric field dynamics:
\begin{align}
    \frac{\partial E_m}{\partial t} = \frac{c^2}{2 \pi \chi \rho} \nabla^2 E_m - \frac{1}{2 \pi} E_m \nabla \cdot {\bm v}.
\end{align}
We find that there is an effective diffusion of the fluctuations in the electric field and a nonlinear forcing due to the presence of compressible velocity structures. While in MHD there is a turbulent dynamo action this translates to the effective dynamo action in the hydrodynamics of the electric field where there is a spontaneous generation of electric field fluctuations in compression fronts where $\nabla \cdot {\bm v} < 0$. Now the Navier-Stokes equation has the form:
\begin{align}
    \rho_0 \frac{\partial {\bm v}}{\partial t} = \eta \nabla^2 {\bm v} + \eta^o \nabla^2 ({\bm \epsilon} \cdot {\bm v}) - a \nabla \rho - \frac{1}{\rho \pi} E_m \nabla E_m
\end{align}
 We now linearize the system about an electric field $E_0$ and obtain:
\begin{subequations}
\begin{equation}
     \frac{\partial \rho}{\partial t} = -\rho_0 \nabla \cdot {\bm v}
     \end{equation}
\begin{equation}
     \rho_0 \frac{\partial {\bm v}}{\partial t} = \eta \nabla^2 {\bm v} + \eta^o \nabla^2 ({\bm \epsilon} \cdot {\bm v}) - a \nabla \rho - \frac{1}{\pi} E_0 \nabla E_m, \end{equation}
\begin{equation}
     \frac{\partial E_m}{\partial t} = D \nabla^2 E_m - \frac{1}{2 \pi} E_0 \nabla \cdot {\bm v},
\end{equation}
\end{subequations}
where, diffusivity $D=\frac{c^2}{2 \pi \chi \rho} $. We obtain the dispersion relation (for $\eta=0$ and $D=0$):
\begin{align}
    & \omega = \pm k \sqrt{a + \frac{E_0^2}{2 \pi^2 \rho_0} + \frac{{\eta^o}^2 k^2}{\rho_0^2}}.
\end{align}
We find in the above dispersion relation the effect of density fluctuations, odd viscosity, and the mean electric field. This form is analogous to the magnetosonic waves in MHD systems. Now, we perform the same calculations in the limit of nonzero diffusivity and viscosity. Since the expressions become rather cumbersome we expand in powers of $k$ and retain the leading order terms. We obtain diffusive modes:
\begin{align}
    & \omega_1  = i \frac{2 a D \rho_0 \pi^2 }{E_0^2 + 2 a \pi^2 \rho_0} k^2 + \mathcal{O} (k^4), \nonumber \\
    & \omega_2 = i \frac{\eta}{\rho_0} k^2 + \mathcal{O} (k^4),
\end{align}
and damped waves given by :
\begin{widetext}
\begin{align}
     \omega_{3,4} = & \pm  k \sqrt{\frac{E_0^2 + 2 a \pi^2 \rho_0}{2 \pi^2 \rho_0}} + i k^2 \frac{2 a \eta \pi^2 \rho_0 + E_0^2 (\eta + D \rho_0)}{2 \rho_0 (E_0^2+ 2 a \pi^2 \rho_0)} \nonumber \\
    & \pm k^3 \frac{\pi \rho_0 (4 a^2 (\eta^2 - 4 {\eta^o}^2) \pi^4 \rho_0^2 + E_0^4 (\eta^2 - 4 {\eta^o}^2 - 2 D \eta \rho_0 + D^2 \rho_0^2) + 4 a E_0^2 \pi^2 \rho_0 (\eta^2 - 4 {\eta^o}^2 - D \eta \rho_0 + 2 D^2 \rho_0^2))}{4 \sqrt{2} (\rho_0 (E_0^2 + 2 a \pi^2 \rho_0))^{5/2}} + \mathcal{O} (k^4).
\end{align}
\end{widetext}
In the above expressions $\omega_{1,2,3,4}$ are the expressions for frequency $\omega$. 

\section{Appendix C : Quasi two-dimensional MHD} \label{app:C}

The MHD equations that we are going to consider now can be written as a Lorentz force in the momentum conservation equation coupled to Maxwell's equations giving the evolution of the magnetic field. We first consider Ohm's law in the form:
\begin{equation}
    {\bm E} + {\bm v} \times {\bm B} = D {\bm J} + d_i {\bm J} \times {\bm B} ,
\end{equation}
where, ${\bm E}$ is the electric field, ${\bm v}$ is the velocity field, ${\bm B}$ is the magnetic field, ${\bm J}$ is the current density field, $D$ is the resistivity, and $d_i$ is the ion inertial length. The presence of $d_i$ ensures that Hall effect of the electrons in the presence of magnetic field is taken into account in the Ohm's law. The presence of the ion inertial length in these equations makes them what is called Hall MHD equations~\cite{Banerjee2013,Yadav2021,Mininni2003,Mininni2007,Shaikh2009,Galtier2007,Sahraoui2007, Meyrand2012}. By taking a curl of the above equation and using the Maxwell's equation $\partial_t {\bm B} = - c \nabla \times {\bm E}$ we can eliminate the induced electric field and write the equation in terms of ${\bm B}$ and ${\bm J}$ as:
\begin{equation}
    \frac{\partial}{\partial t} {\bm B} = \nabla \times ({\bm v} \times {\bm B}) - d_i \nabla \times ({\bm J} \times {\bm B}) - D \nabla \times {\bm J}.
\end{equation}

We now further note that in the dynamics of plasma the time derivative of the electric field has negligible contribution which allows us the use the Maxwell's equation $(4 \pi / c){\bm J} = \nabla \times {\bm B}$. Finally, the momentum conservation is given by the Navier-Stokes equation where the magnetic field enters as the Lorentz force term:
\begin{subequations}
\begin{equation}
    \frac{\partial}{\partial t} (\rho {\bm v}) +  \nabla \cdot ( \rho {\bm v} \otimes {\bm v}) = \eta \nabla^2 {\bm v} + \eta^o \nabla^2 {\bm v}^* - \nabla p + \rho {\bm J} \times {\bm B},      \end{equation}
\begin{equation}
     \frac{\partial}{\partial t} {\bm B} = \nabla \times \left[ ({\bm v} - d_i {\bm J}) \times {\bm B} \right] + D \nabla^2 {\bm B},      \end{equation}
\begin{equation}
     \frac{\partial}{\partial t} \rho = - \nabla \cdot (\rho {\bm v}),
\end{equation}
\end{subequations}
where, $\rho$ is the mass density field, $p$ is the hydrostatic pressure, $\eta$ is the viscosity, ${\bm v}^* = {\bm \epsilon} \cdot {\bm v}$, and ${\bm \epsilon}$ is the two dimensional Levi-Civita symbol. While, writing the above equations we have implicitly assumed a two dimensional problem where we can write an unique odd viscosity with the coefficient $\eta^o$. Another important thing to do before we can close the equations is to use an equation of state for $p$. In principle pressure can be a complicated function of temperature, internal energy, density e.t.c. However, we will use a barotropic pressure i.e. pressure depends on density only. This approximation allows us to close the equations without writing additional equations for other variables. Examples of barotropic equation of state are the isothermal equation of state $p = c_1 \rho$ and the adiabatic equation of state $p = c_2 \rho^\gamma$. We will mention our choice of equation of state in the section below where we use it. 

When we are investigating a quasi-two dimensional system then we find that the Hall effect term also gets dropped. To see this let us consider the nature of the term i.e. $\nabla \times ({\bm J} \times {\bm B})$. Now, ${\bm J}$ is directed along the third dimension i.e. along $\hat{\bm z}$ and ${\bm B}$ is in the $x$--$y$ plane. The cross product ${\bm J} \times {\bm B}$ then lies in the $x$--$y$ plane and therefore its' curl is directed along the $z$ axis. Therefore, in the ensuing discussion in this paper we drop the Hall effect term. However, if we are considering a plasma which has been effectively two-dimensionalized due to the presence of a strong magnetic field in the $z$ direction then the Hall effect remains. Another important simplification can be done to the induction equation to incorporate the constraint $\nabla \cdot {\bm B} = 0$. Since the divergence is zero the curl must be enough to define the magnetic field (modulo gauge fixing). Therefore, the induction equation reduces to:
\begin{align}
    \frac{\partial}{\partial t} J = \nabla \times \nabla \times \left( {\bm v} \times {\bm B} \right) + D \nabla^2 J
\end{align}
where, $J$ is pseudo-scalar variable in two dimensions and is defined as ${\bm J} = J \hat{z}$.

Let us now consider the above equations linearized about a mean magnetic field given by $b_0 \hat{x}$. We the obtain the linear system of equations given by:
\begin{subequations}
\begin{equation}
    \frac{\partial}{\partial t} {\bm v} = \nu \nabla^2 {\bm v} + \nu^o \nabla^2 {\bm v}^* - a \nabla \rho + b_0 J \hat{y}      \end{equation}
\begin{equation}
    \frac{\partial}{\partial t} J = b_0 \nabla^2 v_y + D \nabla^2 J      \end{equation}
\begin{equation}
    \frac{\partial}{\partial t} \rho = -\rho_0 \nabla \cdot {\bm v}
\end{equation}
\end{subequations}
where, we have used $\eta/\rho_0 = \nu$ and $\eta^o/\rho_0 = \nu^o$. Here, we have also used an equation of state such that $\nabla p/ \rho = a \nabla \rho$. Also, the velocity coupling in the induction equation reduces to $-b_0 k^2 v_y$ when written in a Fourier space notation. We can write the equations in the form of matrix equations of the form shown in Eq.~\ref{Eq:matrix}.
\begin{widetext}
\begin{align}
\label{Eq:matrix}
      -i \omega \begin{pmatrix}  
                    \rho \\
                    v_x \\
                    v_y \\
                    j
              \end{pmatrix} =
           & \begin{pmatrix}
                    0 & -i \rho_0 k \cos \theta & -i \rho_0 k \sin \theta & 0 \\
                    -i a k \cos \theta & -\nu k^2 & -\nu^o k^2 & 0 \\
                    -i a k \sin \theta & \nu^o k^2 & -\nu k^2 & b_0 \\
                    0 & 0 & -b_0 k^2 & -D k^2 
              \end{pmatrix}
              \begin{pmatrix}  
                    \rho \\
                    v_x \\
                    v_y \\
                    j
              \end{pmatrix}
\end{align}
\end{widetext}

For the purpose of simplicity we neglect $D$ and $\nu$ in the calculations of dispersion relation and we obtain dispersion relations given in Eq.~\ref{Eq:dis1}.
\begin{widetext}
\begin{align}
\label{Eq:dis1}
    \omega = & \pm k \left( \frac{b_0^2}{2} + \frac{k^2 {\nu^o}^2}{2} + \frac{a \rho_0}{2} \pm \frac{1}{2} \sqrt{ b_0^4 + 2 b_0^2 k^2 {\nu^o}^2 + (k^2 {\nu^o}^2 + a \rho_0)^2 - 2 a b_0^2 \rho_0 \cos {2\theta}}\right)^{1/2}.
\end{align}
\end{widetext}

The above dispersion relations give the slow and fast magnetosonic waves characteristic of compressible ($\nabla \cdot {\bm v} \neq 0$) ionic plasma with the contribution from odd viscosity. Typically, magnetosonic waves have contributions from the Alfv\'en waves ($v_a^2 \equiv b_0^2 $) and sound ($v_s^2 \equiv a \rho_0$). In the presence of odd viscosity we find an additional scale dependent velocity which we call $v_o^2 \equiv {\nu^o}^2 k^2$ which becomes relevant for large wave numbers i.e. small length scales. Note, that $\theta$ is a polar angle defined as $\tan \theta = k_y/k_x$ (more generally, $\tan \theta = k_{\perp}/ k_{\parallel}$).\footnote{Note that for incompressible fluids i.e. if $\nabla \cdot {\bm v} = 0$ both the sound waves and the odd waves disappear and we are left with only the Alfv\'en waves. } To further simplify the computation we can write the above dispersion relation as given by
\begin{widetext}
\begin{align}
\label{Eq:dis2}
    2 \frac{\omega^2}{k^2} &= (b_0^2 + k^2 {\nu^o}^2 + a \rho_0) \left( 1 \pm \sqrt{1 - \frac{4 a b_0^2 \rho_0}{(b_0^2 + k^2 {\nu^o}^2 + a \rho_0)^2} \cos^2 \theta} \right) \nonumber \\
    &= \left( v_a^2 + v_o^2 + v_s^2 \right) \left( 1 \pm \sqrt{1 - \frac{4 v_a^2 v_s^2}{(v_a^2 + v_o^2 + v_s^2)^2} \cos^2 \theta}\right)
\end{align}
\end{widetext}

In the above dispersion relation if we consider the specific case of longitudinal waves when ${\bf k} \parallel b_0 \hat{x}$ i.e. $\theta = 0$ we obtain $\omega = \pm k \sqrt{b_0^2 + k^2 {\nu^o}^2}$ and $\omega = \pm k \sqrt{a \rho_0}$. Therefore, the longitudinal Alfv\'en waves are modified in the presence of odd viscosity. In the transverse modes, i.e. ${\bf k} \perp b_0 \hat{x}$ and $\theta = \pi/2$ we obtain $\omega = \pm k \sqrt{b_0^2 + k^2 {\nu^o}^2 + a \rho_0}$. In the limit of weak odd viscosity contribution i.e. $({\nu^o}^2 k^2)/b_0^2 \rightarrow 0 $ we can expand the above dispersion relations and obtain the relations $\omega = \pm k b_0 (1 + {\nu^o}^2 k^2/2 b_0^2 + \mathcal{O} (k^4))$ for longitudinal modes and $\omega = \pm k (b_0 + \sqrt{a \rho_0}) (1 + {\nu^o}^2 k^2/2 (b_0^2 + a \rho_0) + \mathcal{O} (k^4) )$ for the transverse waves. We find in the above limit the emergence of dispersive waves with $\omega \sim k^3$ corrections due to odd viscosity. Such a ($k^3$) dispersion is observed in the one dimensional KdV equations.

\end{document}